\documentclass[12pt]{article}
\usepackage{graphicx,amssymb,amsfonts,amsmath,epsfig}

\makeatletter
\DeclareSymbolFont{AMSb}{U}{msb}{m}{n}
\DeclareSymbolFontAlphabet{\mathbb}{AMSb}
\renewcommand{\section}{\@startsection{section}{1}{\z@}%
                                    {-7ex \@plus -1ex \@minus -.2ex}%
                                    {2.5ex \@plus.2ex}%
                                    {\normalfont\large\scshape\centering}}
\renewcommand{\subsection}{\@startsection{subsection}{2}{\z@}%
                                       {-5ex \@plus -1ex \@minus -.2ex}%
                                       {1.5ex \@plus.2ex}%
                                       {\normalfont\normalsize\scshape}}
\renewcommand{\subsubsection}{\@startsection{subsubsection}{3}{\z@}%
                                       {-5ex \@plus -1ex \@minus -.2ex}%
                                       {1.5ex \@plus.2ex}%
                                       {\normalfont\normalsize\scshape}}

\renewcommand\@seccntformat[1]{\ignorespaces\csname #1name\endcsname\space
                               \csname the#1\endcsname.\quad}   
\newdimen\captionmargin
\newdimen\captionindent
\newdimen\captionwidth
\newcommand{\captionfont}{\slshape}
\newcommand\@captionlabel[1]{\textsc{#1:}\space}
\long\def\@makecaption#1#2{%
  \vskip\abovecaptionskip
  \captionwidth\hsize
  \advance\captionwidth -2\captionmargin
  \sbox\@tempboxa{\@captionlabel{#1}\captionfont #2}%
  \ifdim \wd\@tempboxa >\captionwidth
    \ifdim\captionindent>\z@
      \advance\captionwidth -\captionindent
      \hskip\captionindent
    \fi
    \hskip\captionmargin
    \parbox[t]{\captionwidth}{\leavevmode\hskip-\captionindent
      \@captionlabel{#1}\captionfont #2}%
  \else
    \global \@minipagefalse
    \hb@xt@\hsize{\hfil\box\@tempboxa\hfil}%
  \fi
  \vskip\belowcaptionskip}
\def\eqnarray{%
   \stepcounter{equation}%
   \def\@currentlabel{\p@equation\theequation}%
   \global\@eqnswtrue
   \m@th
   \global\@eqcnt\z@
   \tabskip\@centering
   \let\\\@eqncr
   $$\everycr{}\halign to\displaywidth\bgroup
       \hskip\@centering$\displaystyle\tabskip\z@skip{##}$\@eqnsel
      &\global\@eqcnt\@ne$\;\hfil{##}$\hfil
      &\global\@eqcnt\tw@$\;\displaystyle{##}$\hfil\tabskip\@centering
      &\global\@eqcnt\thr@@ \hb@xt@\z@\bgroup\hss##\egroup
         \tabskip\z@skip
      \cr}
\ifcase \@ptsize
\or
\or
\fi
  \divide\@tempdima\baselineskip
  \@tempcnta=\@tempdima
\makeatother
\begin{document}

%
%

\renewcommand{\theequation}{\arabic{section}.\arabic{equation}}
\renewcommand{\thefigure}{\arabic{figure}}
\newcommand{\gapprox}{%
\mathrel{%
\setbox0=\hbox{$>$}\raise0.6ex\copy0\kern-\wd0\lower0.65ex\hbox{$\sim$}}}
\textwidth 165mm \textheight 220mm \topmargin 0pt \oddsidemargin 2mm
\def\ib{{\bar \imath}}
\def\jb{{\bar \jmath}}

\newcommand{\ft}[2]{{\textstyle\frac{#1}{#2}}}
\newcommand{\be}{\begin{equation}}
\newcommand{\ee}{\end{equation}}
\newcommand{\bea}{\begin{eqnarray}}
\newcommand{\eea}{\end{eqnarray}}
\newcommand{\Identity}{{1\!\rm l}}
\newcommand{\cx}{\overset{\circ}{x}_2}
\def\CN{$\mathcal{N}$}
\def\CH{$\mathcal{H}$}
\def\hg{\hat{g}}
\newcommand{\bref}[1]{(\ref{#1})}
\def\espai{\;\;\;\;\;\;}
\def\zespai{\;\;\;\;}
\def\avall{\vspace{0.5cm}}
\newtheorem{theorem}{Theorem}
\newtheorem{acknowledgement}{Acknowledgment}
\newtheorem{algorithm}{Algorithm}
\newtheorem{axiom}{Axiom}
\newtheorem{case}{Case}
\newtheorem{claim}{Claim}
\newtheorem{conclusion}{Conclusion}
\newtheorem{condition}{Condition}
\newtheorem{conjecture}{Conjecture}
\newtheorem{corollary}{Corollary}
\newtheorem{criterion}{Criterion}
\newtheorem{defi}{Definition}
\newtheorem{example}{Example}
\newtheorem{exercise}{Exercise}
\newtheorem{lemma}{Lemma}
\newtheorem{notation}{Notation}
\newtheorem{problem}{Problem}
\newtheorem{prop}{Proposition}
\newtheorem{rem}{{\it Remark}}
\newtheorem{solution}{Solution}
\newtheorem{summary}{Summary}
\numberwithin{equation}{section}
\newenvironment{pf}[1][Proof]{\noindent{\it {#1.}} }{\ \rule{0.5em}{0.5em}}
\newenvironment{ex}[1][Example]{\noindent{\it {#1.}}}

\thispagestyle{empty}


\begin{center}

{\LARGE\scshape Emission of fermions in little string theory
\par}
\vskip15mm

\textsc{Oscar Lorente-Esp\'{i}n}
\par\bigskip
{\em
Departament de F{\'\i}sica i Enginyeria Nuclear,
Universitat Polit\`ecnica de Catalunya,\\
Comte Urgell, 187, E-08036 Barcelona, Spain.}\\[.1cm]
\vspace{5mm}
\end{center}

\section*{Abstract}
It is well known that little string theory (LST) black holes radiate a purely thermal spectrum of scalar particles. 
This theory 
lives in a Hagedorn phase with a fixed Hagedorn temperature that does not depend on its mass. Therefore the theory 
keeps a thermal profile even taking into account self-gravitating effects and the backreaction of the metric. 
This has implications concerning the information loss paradox; one would not be able to recover any information from the 
LST black hole since the emission of scalar particles is totally uncorrelated. Several studies of the emission spectrum 
in LST 
concern scalar fields; it is our aim in this work to extend the study to the emission of fermions in order to verify that 
the most relevant conclusion for the scalar field remains valid for the fermion fields. Thus, we have calculated the emission 
probability, the flux and also the greybody factor corresponding to a fermion field in LST background.

\vskip10mm
\noindent KEYWORDS: Black Holes, Hawking radiation, Fermions.

\vspace{3mm} \vfill{ \hrule width 5.cm \vskip 2.mm {\small
\noindent E-mail: oscar.lorente-espin@upc.edu }}



\newpage
\setcounter{page}{1}


\tableofcontents       %
\vskip 1cm             %

\setcounter{equation}{0}

\section{Introduction}
In \cite{Hawking:1974sw} Hawking demonstrated that black holes radiate a purely thermal radiation at a definite temperature.
This fact drove us to the so-called {\it information loss paradox}, which states the violation of quantum unitary evolution.
A lot of work has been done in order to solve this paradox, e.g. great success has been achieved in string theory 
framework using holography principle techniques \cite{Maldacena:1996ky,Peet:2000hn,Das:2000su,Susskind:2005js}.\par

On the other hand, using semiclassical techniques, some authors in 
\cite{Parikh:1999mf,Kraus:1994by,Parikh:2004rh,Arzano:2005rs,Zhang:2009jn} have showed how deviations of the black hole 
radiance from purely a thermal spectrum enable us to recover the information lost. In the work of Hawking it was not 
considered 
the self-gravitation of the emitted particles neither the backreaction of the background metric. Nevertheless, the 
tunneling 
picture proposed in \cite{Parikh:1999mf} takes into account the self-gravitation of the radiation; hence imposing energy 
conservation the metric backreacts when the black hole emits and the event horizon shrinks. The tunneling mechanism 
matches the heuristic picture more considered more closely, namely the creation of a particle-antiparticle pair just 
outside the 
horizon or inside of the black hole. From a pair created outside the horizon one member of the pair, e.g. the antiparticle, 
can tunnels through the horizon falling down the hole whereas the particle escapes out. Concretely this flux is 
detected as Hawking radiation by an asymptotic observer. In the same way, if the pair is 
created inside the horizon, the particle can tunnel through the horizon escaping to the asymptotic, whereas the antiparticle 
falls down to the black hole. Anyway, the flux energy of the incoming antiparticles is negative while the outgoing flux of 
particles is positive. Therefore, taking into account the energy conservation, the total ADM mass is conserved whereas the 
black hole mass decreases by the amount $M\rightarrow M-\omega$, where $\omega$ is the energy of the emitted particle. As 
the event 
horizon radius is proportional to the mass, the black hole will be shrunk. It is noteworthy to say that in this tunneling 
mechanism is the emitted particle which creates the potential barrier \cite{Parikh:2004rh}.\par  

In \cite{LorenteEspin:2007gz,LorenteEspin:2011pu} we studied the Hawking radiation for NS5 and little string theory (LST) 
black holes using different semiclassical methods, i.e. the tunneling and the complex path method. 
We verified that the NS5 model showed a nonthermal profile whereas LST showed a thermal behavior. This last conclusion 
matches the 
Hagedorn properties of LST, namely the temperature of LST corresponds to the Hagedorn temperature. However, we only studied
the emission of a scalar field. It would be interesting to extend the study of the Hawking emission by LST to 
fermion fields and verify if the LST spectrum changes or remains purely thermal. 
\par
The paper is organized as follows. In Sec. II, we briefly summarize some properties and thermodynamics of LST. 
In Sec. III, we solve the Dirac equation for a fermion field in LST background. Next, we study the emission probability of 
fermions. Also, we calculate the spectrum and the radiation flux corresponding to a massless fermion field, and we 
analyze the implications of considering the energy conservation and the backreaction of the metric when 
the black hole emits. Finally, in Sec. IV, we discuss the most relevant results.

\section{Description of the model}
LST \footnote{For good reviews on LST and NS5 we address the readers to 
\cite{Kutasov:2001uf, Aharony:1999ks, Kapustin:1999ci, Kutasov:2000jp, Rangamani:2001ir, Aharony:1998ub}.} 
is a nongravitational six dimensional and nonlocal field theory 
\cite{Kutasov:2001uf, Aharony:1999ks, Kapustin:1999ci, Peet:1998wn, Seiberg:1997zk}, believed 
to be dual to a string theory background, defined as the decoupled theory on a stack of N NS5-branes. In the 
limit of a vanishing asymptotic value for the string coupling $g_{s}\rightarrow0$, keeping the string length $l_{s}$ 
fixed while the energy above extremality is fixed, i.e. $\frac{E}{m_{s}}={\rm fixed}$, the processes in which the 
modes that live on 
the branes are emitted into the bulk as closed strings are suppressed. The theory becomes free in the bulk, but 
strongly interacting on the brane. In this limit, the theory reduces to the little string theory or more precisely to 
(2,0) LST for type IIA NS5-branes and to (1,1) LST for type IIB NS5-branes \cite{Aharony:1998ub}.\par
The throat geometry corresponding to N coincident nonextremal NS5-branes in the string frame \cite{Maldacena:1997cg} is

\begin{equation}
 \label{metric}
ds^{2}=-f(r)dt^2+\frac{A(r)}{f(r)}dr^2+A(r)r^2d\Omega_{3}^2+\sum_{j=1}^{5}dx_{j}^2\;,
\end{equation}
where $dx_{j}^2$ corresponds to flat spatial directions along the 5-branes, $d\Omega_{3}^2$ corresponds to 3-sphere of
the transverse geometry and the dilaton field is defined as 
\begin{equation}
 \label{dilaton}
e^{2\Phi}=g_{s}^{2}A(r) \;.
\end{equation}
The metric functions are 
\begin{equation}
 \label{metricf}
f(r)=1-\frac{r_{0}^2}{r^2}\;\;\;,\;\;\;A(r)=\chi+\frac{N}{m_{s}^2r^2}\;,
\end{equation}
where $r_{0}$ is 
the location of the event horizon. We define the parameter $\chi$ which takes the values 
1 for NS5 model and 0 for LST; these are only the values for which exists a supergravity solution. In addition to the 
previous fields, there 
is a $NS-NS$ $H_{(3)}$ form along the $S^3$, $H_{(3)}=2N\epsilon_{3}$. According to the holographic principle, the high
spectrum of this dual string theory should be approximated by a certain black hole in the background (\ref{metric}). 
The geometry transverse to the 5-branes is a long
tube which opens up into the asymptotic flat space with the horizon at the other end. In the limit $r\rightarrow r_{0}$
appears the semi-infinite throat parametrized by (t,r) coordinates. In this region the dilaton grows linearly pointing out 
that gravity becomes strongly coupled far down the throat. The string propagation in this 
geometry should correspond to an exact conformal field theory \cite{Kutasov:1990ua}. The boundary of the near horizon 
geometry is $R^{5,1}\times R\times S^{3}$.

\section{Tunneling of fermions in LST}
In this section we will study the tunneling of fermions through the event horizon of the LST background (\ref{metric}). 
We write the covariant Dirac equation in a general background \cite{Brill:1957fx} for a spinor field $\Psi$,
\begin{equation}
 \label{04dirac1}
\left(\gamma^{a}e_{a}^{\mu}(\partial_{\mu}+\Gamma_{\mu})+m\right)\Psi=0\;,
\end{equation}
where $m$ is the bare mass of the particle; $e_{a}^{\mu}$ are the vielbein defined by the relation 
$g_{\mu\nu}=\eta_{ab}\;e_{\mu}^{a}e_{\nu}^{b}$ with $\eta_{ab}={\rm diag}(-1,1,1,1,...)$; and the latin indexes run for 
local inertial flat coordinates $(0,1,2,...)$, whereas the greek indexes run for general coordinates $(t,r,\theta,...)$. 
In LST, the vielbein take the form
\begin{equation}
 \label{04vielbeins}
e_{\mu}^{a}={\rm diag}\left(\sqrt{f(r)},\sqrt{\frac{A(r)}{f(r)}},r\sqrt{A(r)},r\sqrt{A(r)}\sin\theta,
r\sqrt{A(r)}\sin\theta\sin\varphi,1,...,1\right)\;.
\end{equation}
The spin connection is defined as
\begin{equation}
 \label{04connection}
\Gamma_{\mu}=\frac{1}{8}[\gamma^{c},\gamma^{b}]e_{c}^{\nu}\nabla_{\mu}e_{b\nu}\;,
\end{equation}
where $\nabla_{\mu}e_{b\nu}=\partial_{\mu}e_{b\nu}-\Gamma_{\mu\nu}^{\lambda}e_{b\lambda}$ is the covariant derivative of 
$e_{b\nu}$ and $[\gamma^{c},\gamma^{b}]$ the commutator of the gamma matrices.
We choose for the gamma matrices \cite{Cho:2003qe},
\begin{equation}
 \label{04gammas}
\gamma^{0} = \left(
\begin{array}{ccc}
-i & 0 \\
 0 & i
\end{array} \right)\;\;\;,\;\;\;
\gamma^{k} = \left(
\begin{array}{ccc}
                0 & -i\sigma^{k} \\
      i\sigma^{k} & 0
\end{array} \right)\;\;\;,\;\;\;
k=1,2,3\;;
\end{equation}
where the matrices $\sigma^{i}$ are the Pauli matrices, 
\begin{equation}
 \label{04paulimatrices}
\sigma^{1} = \left(
\begin{array}{ccc}
 0 & 1 \\
 1 & 0
\end{array} \right)\;\;\;,\;\;\;
\sigma^{2} = \left(
\begin{array}{ccc}
 0 & -i \\
 i & 0
\end{array} \right)\;\;\;,\;\;\;
\sigma^{3} = \left(
\begin{array}{ccc}
 1 & 0 \\
 0 & -1
\end{array} \right) \;.
\end{equation}
The gamma matrices satisfies the Clifford algebra,
\begin{eqnarray}
 \label{04clifford}
&& [\gamma^{a},\gamma^{b}]=-[\gamma^{b},\gamma^{a}]\;\;\;{\rm if\;\;a\neq b}, \nonumber\\
&& [\gamma^{a},\gamma^{b}]=0\;\;\;{\rm if\;\;a=b}\;, \nonumber\\
&& \{\gamma^{\mu},\gamma^{\nu}\}=2g^{\mu\nu}\;, \nonumber\\
&& \{\gamma^{a},\gamma^{b}\}=2\eta^{ab} \;.
\end{eqnarray}
Thus, taking into account all the aforesaid definitions and properties, we write the full ten-dimensional Dirac equation for 
the spinor field in the LST background,
\begin{equation}
 \label{04dirac10}
\begin{split}
& \left(\gamma^{0}\frac{1}{\sqrt{f(r)}}\;\partial_{t}+
\gamma^{1}\left(\frac{3A(r)'\sqrt{f(r)}}{4A(r)^{\frac{3}{2}}}+
\frac{6\sqrt{f(r)}}{4r\sqrt{A(r)}}+\frac{f(r)'}{4\sqrt{A(r)f(r)}}+
\sqrt{\frac{f(r)}{A(r)}}\;\partial_{r}\right)+ \right.\\
& +\gamma^{2}\frac{1}{r\sqrt{A(r)}}(\cot\theta+\partial_{\theta})+
\gamma^{3}\frac{1}{r\sqrt{A(r)}}\sin\theta\;\partial_{\varphi}+
\gamma^{4}\frac{1}{r\sqrt{A(r)}}\sin\theta \sin\varphi\;\partial_{\psi}+ \\
& +\left. \sum_{j=5}^{9}\gamma^{j}\partial_{x_{j}}+m\bigg)\Psi=0 \;,
\right.\end{split}
\end{equation}
where prime denotes derivative with respect to the $r$ coordinate.

\subsection{Emission probability}
Henceforth, in order to study the probability emission of fermions, we will be interested in the $r-t$  sector of the LST 
metric, \footnote{In \cite{LorenteEspin:2011pu}, the reduction from the ten-dimensional to the two-dimensional 
metric of LST is shown.} 
\begin{equation}
 \label{effmetric}
ds_{eff}^2=-f(r)dt^2+\frac{A(r)}{f(r)}dr^2 \;.
\end{equation}
Working with this effective two-dimensional metric, the Dirac equation corresponding to a spinor field is simplified to
\begin{equation}
 \label{04dirac2}
\left(\gamma^{0}\frac{1}{\sqrt{f(r)}}\;\partial_{t}+
\gamma^{1}\left(\frac{f(r)'}{4\sqrt{A(r)f(r)}}+
\sqrt{\frac{f(r)}{A(r)}}\partial_{r}\right)+m\right)\Psi=0 \;.
\end{equation}
Taking into account the appropriate choice of the gamma matrices (\ref{04gammas}), we use for the spin-up and spin-down 
Dirac fields, respectively, the following Wentzel-Kramers-Brillouin (WKB) ansatz \cite{Kerner:2007rr}:
\begin{equation}
 \label{04spinorup}
\Psi_{\uparrow}=\left(
\begin{array}{c}
{\cal A}(t,r)\xi_{\uparrow} \\
0 \\
0 \\
{\cal D}(t,r)\xi_{\uparrow}
\end{array} \right)
\exp\left[\frac{i}{\hbar}S_{\uparrow}(t,r)\right]\;,
\end{equation}
\begin{equation}
 \label{04spinordown}
\Psi_{\downarrow}=\left(
\begin{array}{c}
0 \\
{\cal B}(t,r)\xi_{\downarrow} \\
{\cal C}(t,r)\xi_{\downarrow} \\
0
\end{array} \right)
\exp\left[\frac{i}{\hbar}S_{\downarrow}(t,r)\right]\;,
\end{equation}
where $S$ is the classical action, whereas ${\cal A}$, ${\cal B}$, ${\cal C}$ and ${\cal D}$ are arbitrary functions of the 
coordinates. Measuring the spin in the $z$ direction, the eigenvector of $\sigma^{3}$ for the spin-up and spin-down fields, 
respectively, are $\xi_{\uparrow}=\left(
\begin{array}{ccc}
 1 \\
 0
\end{array} \right)$ and 
$\xi_{\downarrow}=\left(
\begin{array}{ccc}
 0 \\
 1
\end{array} \right)$.
We will only solve the spin-up case and the spin-down case is solved analogously. Thus, we substitute the spinor-up field 
(\ref{04spinorup}) and the gamma matrices $\gamma^{0}$ and $\gamma^{1}$ into the Dirac equation (\ref{04dirac2}). Next, we 
apply the WKB approximation neglecting the $\hbar$ dependent terms, and after some algebra 
we eventually obtain the following set of equations for the spin-up case:
\begin{equation}
 \label{04wkbdirac2up}
\begin{array}{ccc}
\left(-\frac{1}{\sqrt{f(r)}}\;\partial_{t}S_{\uparrow}(t,r)+m\right){\cal A}(t,r)-
\sqrt{\frac{f(r)}{A(r)}}\;\partial_{r}S_{\uparrow}(t,r)\;{\cal D}(t,r)=0 \;, \cr
& \cr
\sqrt{\frac{f(r)}{A(r)}}\;\partial_{r}S_{\uparrow}(t,r)\;{\cal A}(t,r)
+\left(\frac{1}{\sqrt{f(r)}}\;\partial_{t}S_{\uparrow}(t,r)+m\right){\cal D}(t,r)=0 \;.
\end{array}
\end{equation}
For the spin-down case, we would obtain
\begin{equation}
 \label{04wkbdirac2down}
\begin{array}{ccc}
\left(-\frac{1}{\sqrt{f(r)}}\;\partial_{t}S_{\downarrow}(t,r)+m\right){\cal B}(t,r)-
\sqrt{\frac{f(r)}{A(r)}}\;\partial_{r}S_{\downarrow}(t,r)\;{\cal C}(t,r)=0 \;, \cr
& \cr
\sqrt{\frac{f(r)}{A(r)}}\;\partial_{r}S_{\downarrow}(t,r)\;{\cal B}(t,r)
+\left(\frac{1}{\sqrt{f(r)}}\;\partial_{t}S_{\downarrow}(t,r)+m\right){\cal C}(t,r)=0 \;.
\end{array} 
\end{equation}
In order to obtain nonvanishing values of the functions ${\cal A}$, ${\cal B}$, ${\cal C}$ and ${\cal D}$, 
(\ref{04wkbdirac2up}) must fulfill the following condition:
\begin{equation}
 \label{04condition2}
\left|
 \begin{array}{cc}
 -\frac{1}{\sqrt{f(r)}}\;\partial_{t}S_{\uparrow}(t,r)+m & 
 -\sqrt{\frac{f(r)}{A(r)}}\;\partial_{r}S_{\uparrow}(t,r) \\
 \sqrt{\frac{f(r)}{A(r)}}\;\partial_{r}S_{\uparrow}(t,r) & 
\frac{1}{\sqrt{f(r)}}\;\partial_{t}S_{\uparrow}(t,r)+m
 \end{array}
\right|=0 \;.
\end{equation}
Writing the action as an expansion in a power series of $(\frac{\hbar}{i})$, 
\begin{equation}
 \label{04actionexpansion}
S_{\uparrow}(t,r)=S_{0\uparrow}(t,r)+\left(\frac{\hbar}{i}\right)S_{1\uparrow}(t,r)+
\left(\frac{\hbar}{i}\right)^2S_{2\uparrow}(t,r)+...
\end{equation}
and making use of the WKB approximation (we neglect terms of order $(\frac{\hbar}{i})$ and higher); we finally 
obtain a nonlinear first-order partial differential equation which corresponds to the Hamilton-Jacobi 
equation of motion to the leading order in the action $S_{\uparrow}$,
\begin{equation}
\label{04hjeom2}
-A(r)\left(\frac{\partial S_{0\uparrow}(t,r)}{\partial t}\right)^2+
f(r)^{2}\left(\frac{\partial S_{0\uparrow}(t,r)}{\partial r}\right)^2+
A(r)f(r)m^{2}=0\;.
\end{equation}

Previously, in \cite{LorenteEspin:2011pu}, it was obtained the same Hamilton-Jacobi equation corresponding, in that case, 
to the propagation of a massless scalar field. The leading-order action solution is
\begin{equation}
 \label{actionmassless}
S_{0\uparrow}(t,r)=-\omega t\pm\omega\int_{r_{in}}^{r_{out}}\frac{\sqrt{A(r)}}{f(r)}\;dr \;.
\end{equation}
The plus/minus sign corresponds to ingoing/outgoing fermions, respectively; $r_{in}$ and $r_{out}$ corresponds to a position 
inside and outside of the black hole, respectively; and $\omega$ is the energy of the emitted or absorbed fermion. 
Then, performing the complex integration of (\ref{actionmassless}) and using the saddle point approximation for the 
amplitude, $K(r_{out},t_{2};r_{in},t_{1})=N\exp\left[\frac{i}{\hbar}S_{0}(r_{out},t_{2};r_{in},t_{1})\right]$, we
obtain the tunneling emission probability,
\begin{equation}
 \label{emprob2}
P_{e}\sim\exp\left[-\frac{\pi}{\hbar}\omega\sqrt{N}\right] \;,
\end{equation}
where we can identify the Hagedorn temperature of LST as 
\begin{equation}
 \label{temp}
T_{H}=\frac{\hbar m_{s}}{2\pi\sqrt{N}} \;,
\end{equation}
$m_{s}$ being the string mass.\par
We have found that the same results for the massless scalar particles in 
\cite{LorenteEspin:2011pu} are obtained here for fermions. So, we can conclude that 
the tunneling emission through the event horizon of the LST does not depend on the particle's spin.\par

We would like to consider briefly the case for massive particles. If we solve (\ref{04hjeom2}) by taking into account the  
mass term, we obtain the following action at leading order:
\begin{equation}
 \label{04action2}
S_{0\uparrow}(t,r)=-\omega t\pm\int_{r_{in}}^{r_{out}}\frac{\sqrt{A(r)}}{f(r)}\;\sqrt{\omega^2-f(r)m^2}\;dr \;.
\end{equation}
We notice that one obtains again the same results for the emission probability and temperature, thus the emission does not 
depend on the mass of the emitted particle.

\subsection{Fermion modes and backscattering spectrum}
We are interested in calculating the average number of emitted fermions as well as the radiation flux. 
In \cite{Banerjee:2009wb}, the authors calculated the Hawking blackbody spectrum corresponding to a spherically symmetric 
static black hole. It is our aim to perform a similar analysis corresponding to LST black hole. 
Our starting point is the two-dimensional action (\ref{actionmassless}), which can be written as 
\begin{equation}
 \label{034action2}
S_{0}(r,t)=\omega(t\pm r^{*}) \;,
\end{equation}
where $r^{*}$ is the tortoise coordinate defined as 
\begin{equation}
 \label{034tortoise}
dr^{*}=\frac{\sqrt{A(r)}}{f(r)}dr \;.
\end{equation}
Then if we consider the outgoing/ingoing null coordinates
\begin{equation}
 \label{034nullcoordinates}
u=t-r^{*}\;\;\;,\;\;\;v=t+r^{*} \;,
\end{equation}
we can define the right/left modes inside and outside of the black hole in the following way:
\begin{eqnarray}
 \label{034modes}
&& \phi^{R}_{in} = e^{-\frac{i}{\hbar}\omega u_{in}}\;\;,\;\;\phi^{L}_{in} = e^{-\frac{i}{\hbar}\omega v_{in}}, \nonumber\\
&& \phi^{R}_{out} = e^{-\frac{i}{\hbar}\omega u_{out}}\;\;,\;\;\phi^{L}_{out} = e^{-\frac{i}{\hbar}\omega v_{out}} \;.
\end{eqnarray}
The Kruskal coordinates, see e.g. \cite{Misner:1974qy}, corresponding to the inside and outside of the LST black 
hole are defined as
\begin{eqnarray}
 \label{034kruskal}
&& T_{in}=e^{\kappa r^{*}_{in}}\cosh(\kappa t_{in})\;\;,\;\;X_{in}=e^{\kappa r^{*}_{in}}\sinh(\kappa t_{in}), \nonumber\\
&& T_{out}=e^{\kappa r^{*}_{out}}\sinh(\kappa t_{out})\;\;,\;\;X_{out}=e^{\kappa r^{*}_{out}}\cosh(\kappa t_{out}) \;,
\end{eqnarray}
where $\kappa$ is the surface gravity and is related with the Hawking temperature through the relation
$T_{H}=\frac{\hbar\kappa}{2\pi}$.
The two sets of Kruskal coordinates are then connected
by the following transformation relation between the coordinates $t$ and $r$:
\begin{equation}
 \label{034transformation1}
t_{in}\rightarrow t_{out}-i\frac{\pi}{2\kappa}\;\;,\;\;r^{*}_{in}\rightarrow r^{*}_{out}+i\frac{\pi}{2\kappa} \;.
\end{equation}
Moreover, the null coordinates are also transformed as
\begin{equation}
 \label{034transformation2}
u_{in}\rightarrow u_{out}-i\frac{\pi}{\kappa}\;\;,\;\;v_{in}\rightarrow v_{out}\;.
\end{equation}
Eventually, we have obtained a transformation relation between the left/right modes inside and outside the black hole,
\begin{equation}
 \label{034transformation3}
\phi^{R}_{in}\rightarrow\phi^{R}_{out}\;e^{-\frac{\pi\omega}{\hbar\kappa}}\;\;,\;\;\phi^{L}_{in}\rightarrow\phi^{L}_{out}\;.
\end{equation}
As a comment, we note that this last relation is precisely the relation that one obtains between the Bogoliubov 
coefficients in the standard study of the emission of Hawking radiation.\par

Now, we want to calculate the average number of fermions emitted by the black hole.
We construct the physical state associated to a system of $n$ number of a noninteracting virtual pair of fermions created 
inside the black hole,
\begin{equation}
 \label{e1}
\lvert\psi\rangle=N\sum_{n}\;\lvert n^{L}_{in}\rangle\otimes\lvert n^{R}_{in}\rangle \;.
\end{equation}
Since outside of the black hole we can carry out observations, we want to write the physical state in terms of the out 
eigenstates. Thus using the relation (\ref{034transformation3}) between the modes inside and 
outside of the black hole, we can obtain the desired expression,
\begin{equation}
 \label{e2}
\lvert\psi\rangle=
N\sum_{n}e^{-\frac{\pi\omega n}{\hbar\kappa}}\lvert n^{L}_{out}\rangle\otimes\lvert n^{R}_{out}\rangle\;.
\end{equation}
In order to calculate the normalization constant $N$, we make use of the orthonormalization condition between two
orthonormalized states,
\begin{equation}
 \label{e3}
\langle\psi_{m}\lvert\psi_{n}\rangle=\delta_{mn} \;.
\end{equation}
Then, considering two states $\lvert\psi_{n}\rangle$ and $\langle\psi_{m}\lvert$, we construct
\begin{eqnarray}
 \label{e4}
\langle\psi_{m}\lvert\psi_{n}\rangle & = & 
\left(N\sum_{m}e^{-\frac{\pi\omega m}{\hbar\kappa}}\langle m^{L}_{out}\lvert\otimes\langle m^{R}_{out}\lvert\right)\cdot
\left(N\sum_{n}e^{-\frac{\pi\omega n}{\hbar\kappa}}\lvert n^{L}_{out}\rangle\otimes\lvert n^{R}_{out}\rangle\right) 
\nonumber \\
& = & N^2\sum_{m,n}e^{-\frac{\pi\omega (m+n)}{\hbar\kappa}}\langle m^{L}_{out}\lvert n^{L}_{out}\rangle\otimes
\langle m^{R}_{out}\lvert n^{R}_{out}\rangle \;
\end{eqnarray}
and taking into account (\ref{e3}), we obtain the following relation:
\begin{equation}
 \label{e5}
1=N^2\sum_{n}e^{-\frac{2\pi\omega n}{\hbar\kappa}} \;,
\end{equation}
from which we obtain the normalization constant corresponding to fermions ($n=0,1$),
\begin{equation}
 \label{e6}
N_{f}=\left(1+e^{-\frac{2\pi\omega}{\hbar\kappa}}\right)^{-\frac{1}{2}} \,.
\end{equation} 
A state associated to a system of fermions inside the black hole can be written as
\begin{equation}
 \label{e7}
\lvert\psi_{f}\rangle=\left(1-e^{+\frac{2\pi\omega}{\hbar\kappa}}\right)^{\frac{1}{2}}
\sum_{n}e^{-\frac{\pi\omega n}{\hbar\kappa}}\lvert n^{L}_{out}\rangle\otimes\lvert n^{R}_{out}\rangle\;.
\end{equation}
We might construct the density matrix operator for a system of fermions as
\begin{eqnarray}
 \label{e8}
\rho_{f} & = & \lvert\psi_{n}\rangle\langle\psi_{m}\lvert \nonumber \\
& = & \left(N_{f}\sum_{n}e^{-\frac{\pi\omega n}{\hbar\kappa}}\lvert n^{L}_{out}\rangle\otimes\lvert n^{R}_{out}\rangle
\right)\cdot 
\left(N_{f}\sum_{m}e^{-\frac{\pi\omega m}{\hbar\kappa}}\langle m^{L}_{out}\lvert\otimes\langle m^{R}_{out}\lvert\right)
\nonumber \\
& = & \left(1-e^{-\frac{2\pi\omega}{\hbar\kappa}}\right)
\sum_{n,m}e^{-\frac{\pi\omega (n+m)}{\hbar\kappa}}\left(\lvert n^{L}_{out}\rangle\otimes\lvert n^{R}_{out}\rangle\right)
\cdot
\left(\langle m^{L}_{out}\lvert\otimes\langle m^{R}_{out}\lvert\right) \nonumber \\
& = & \left(1-e^{-\frac{2\pi\omega}{\hbar\kappa}}\right)
\sum_{n,m}e^{-\frac{\pi\omega (n+m)}{\hbar\kappa}}\left(\lvert n^{L}_{out}\rangle\langle m^{L}_{out}\lvert\right)
\otimes\left(\lvert n^{R}_{out}\rangle\langle m^{R}_{out}\lvert\right) \;.
\end{eqnarray}
Tracing over the left modes, 
$\langle m^{L}_{out}\lvert\left(\lvert n^{L}_{out}\rangle\langle m^{L}_{out}\lvert\right)\lvert n^{L}_{out}\rangle$,
and taking into account the orthonormalization condition (\ref{e3}), we obtain
\begin{equation}
 \label{e10}
\rho^{R}_{f}=\left(1-e^{+\frac{2\pi\omega}{\hbar\kappa}}\right)\sum_{n}e^{-\frac{2\pi\omega n}{\hbar\kappa}}
\lvert n^{R}_{out}\rangle\langle n^{R}_{out}\lvert \;.
\end{equation} 
This expression corresponds to the density matrix of fermions in terms of the right outgoing modes. These modes will be 
detected at asymptotic infinity as the Hawking radiation. Finally, using the equation
\begin{eqnarray}
 \label{e11}
\langle n_{f}\rangle & = & Tr(n\cdot\rho^{R}_{f}) \nonumber \\
& = & \left(1-e^{+\frac{2\pi\omega}{\hbar\kappa}}\right)\sum_{n}n\cdot e^{-\frac{2\pi\omega n}{\hbar\kappa}}
\lvert n^{R}_{out}\rangle\langle n^{R}_{out}\lvert \;,
\end{eqnarray}
tracing over the right outgoing modes, 
$\langle m^{R}_{out}\lvert\left(\lvert n^{R}_{out}\rangle\langle n^{R}_{out}\lvert\right)\lvert n^{R}_{out}\rangle$,
and taking into account the orthonormalization condition (\ref{e3}), we obtain the average number of fermions detected at 
asymptotic infinity,
\begin{equation}
 \label{e13}
\langle n_{f}\rangle=\left(1-e^{+\frac{2\pi\omega}{\hbar\kappa}}\right)\sum_{n}n\cdot e^{-\frac{2\pi\omega n}{\hbar\kappa}}
=\frac{1}{e^{\frac{2\pi\omega}{\hbar\kappa}}+1} \;.
\end{equation}

Integrating this last expression over all energy ranges, we obtain the flux of fermions seen by an asymptotic observer,
\begin{equation}
 \label{034flux}
F_{\infty}=\frac{1}{2\pi}\int^{\infty}_{0}\langle n_{f}\rangle\;\omega\;d\omega=
\frac{\hbar^{2}\kappa^{2}}{96\pi}=\frac{\pi}{24}T^{2}_{H} \;,
\end{equation}
where $T_{H}$ is the temperature defined in (\ref{temp}).

It is well known that some emitted radiation will be partially scattered back to the event horizon.
This fact is due to the gravitational potential barrier around the black hole where some fraction of radiation is
reflected back to the hole, acting thus as a filter for the emitted radiation. 
In this way, Eq. (\ref{e13}) must be modified to the following new expression:
\begin{equation}
 \label{e14}
\langle n_{f}\rangle=\frac{\Gamma_{\omega l}}{e^{\frac{2\pi\omega}{\hbar\kappa}}+1} \;,
\end{equation}
where $\Gamma_{\omega l}$ is the greybody factor and it accounts for the deviation from pure Planckian spectrum. 
Henceforth, we will consider the value $\Gamma_{\omega 0}$ since the main contribution to the greybody factor comes from 
the zero angular momentum $l=0$ whenever the relation $\omega M\ll 1$ is fulfilled.  

Greybody factors have a relevant importance because a successful microscopic account of black hole thermodynamics should be 
able to predict them. For example, it is shown in \cite{Das:1996wn} that D-branes provide an account of black hole 
microstates which is successful to predict the greybody factors. There exists a vast literature on how to compute greybody 
factors in the context of the quantum field theory in curved space-time, e.g. 
\cite{Futterman:1988ni, Maldacena:1996ix, Gubser:1996zp, Klebanov:1996gy, Klebanov:1997cx, Gubser:1997yh, Peet:1997es}.\par
Using the solution corresponding to the propagation of a massless scalar field in the background of LST, 
see \cite{Narayan:2001dr}, and applying the usual matching techniques described in the previous references, 
we have verified that the greybody factor is 1. One could expect this value since the pure-thermal behavior characterizes 
the LST emission. Likewise, we expect to obtain identical results for the emission of fermions in LST. Then, we are going 
to find the fermion modes of the Dirac equation (\ref{04dirac2}) corresponding to the two-dimensional LST background 
(\ref{effmetric}), and next using the matching technique we will compute the greybody factor.\par
Now, in order to study the Dirac equation, we choose the following basis for the spinor field:
\begin{equation}
 \label{042spinor}
\Psi(t,r)=\left(
\begin{array}{c}
 \Psi_{+}(t,r) \\
 \Psi_{-}(t,r) 
\end{array} \right) \;,
\end{equation}
considering that each term is a two-component spin-up and spin-down spinor, $\Psi_{+}=\left(
\begin{array}{c}
 \Psi_{+}^{\uparrow} \\
 \Psi_{+}^{\downarrow} 
\end{array} \right)$ and $\Psi_{-}=\left(
\begin{array}{c}
 \Psi_{-}^{\downarrow} \\
 \Psi_{-}^{\uparrow} 
\end{array} \right)$. 
Using this basis for the spinor field and the $\gamma^{0}$ and $\gamma^{1}$ matrices defined in (\ref{04gammas}), we 
obtain two sets of equivalent equations corresponding to the spin-up and spin-down fermion case. We will study the 
spin-up fermion case; equivalently we could also study the spin-down case. Therefore, the Dirac equation becomes
\begin{equation}
 \label{042dirac2d1}
\begin{array}{ccc}
\left(\frac{-i}{\sqrt{f(r)}}\;\partial_{t}+m\right)\Psi_{+}(t,r)-
i\left(\sqrt{\frac{f(r)}{A(r)}}\;\partial_{r}+\frac{f'(r)}{4\sqrt{A(r)f(r)}}\right)
\Psi_{-}(t,r)=0 \;, \cr
& \cr
\left(\frac{i}{\sqrt{f(r)}}\;\partial_{t}+m\right)\Psi_{-}(t,r)+
i\left(\sqrt{\frac{f(r)}{A(r)}}\;\partial_{r}+\frac{f'(r)}{4\sqrt{A(r)f(r)}}\right)
\Psi_{+}(t,r)=0 \;.
\end{array}
\end{equation}
Next, we consider the following ansatz for the spinor field:
\begin{equation}
 \label{042ansatz}
\Psi_{+}(t,r)=\phi_{+}(r)\;e^{-i\omega t}\;\,,\;\;\Psi_{-}(t,r)=i\phi_{-}(r)\;e^{-i\omega t}\;.
\end{equation}
Substituting this expressions into (\ref{042dirac2d1}) and after doing algebra we obtain the following set of 
equations:
\begin{equation}
 \label{042dirac2d2}
\begin{array}{ccc}
\partial_{r}\phi_{-}(r)+\frac{f(r)'}{4f(r)}\;\phi_{-}(r)+\left(m\sqrt{\frac{A(r)}{f(r)}}-
\omega\;\frac{\sqrt{A(r)}}{f(r)}\right)\phi_{+}(r)=0 \;, \cr
& \cr
\partial_{r}\phi_{+}(r)+\frac{f(r)'}{4f(r)}\;\phi_{+}(r)+\left(m\sqrt{\frac{A(r)}{f(r)}}+
\omega\;\frac{\sqrt{A(r)}}{f(r)}\right)\phi_{-}(r)=0 \;.
\end{array}
\end{equation}
This set of coupled equations is analytically solvable. If we define 
\begin{equation}
 \label{042etas}
\eta_{\pm}(r)\equiv m\sqrt{\frac{A(r)}{f(r)}}\pm\omega\;\frac{\sqrt{A(r)}}{f(r)} \;,
\end{equation}
we obtain the following second-order differential equation:
\begin{equation}
 \label{042dirac2d3}
\begin{split}
& \eta_{+}^{-1}(r)\;\phi_{+}(r)''
+\left(\partial_{r}\eta_{+}^{-1}(r)+\eta_{+}^{-1}(r)\;\frac{f(r)'}{2f(r)}\right)\;\phi_{+}(r)'+\\
& +\left(\partial_{r}\eta_{+}^{-1}(r)\;\frac{f(r)'}{4f(r)}+\eta_{+}^{-1}(r)\;\partial_{r}\left(\frac{f(r)'}{4f(r)}\right)
+\eta_{+}^{-1}(r)\left(\frac{f(r)'}{4f(r)}\right)^{2}-\eta_{-}(r)\right)\phi_{+}(r)=0 \;.
\end{split}
\end{equation}
In order to simplify the resolution of the above equations, we consider the propagation of a massless fermion. 
Substituting the values of $f(r)$ and $A(r)$ given in (\ref{metricf}) into (\ref{042etas}) and (\ref{042dirac2d3}), 
eventually we obtain the propagation equation for a massless fermion mode,
\begin{equation}
 \label{042dirac2d4}
4r^{2}(r^{2}-r_{0}^{2})^{2}\;\phi_{+}(r)''+4r(r^{2}-r_{0}^{2})(r^{2}+2r_{0}^{2})\;\phi_{+}(r)'
+(4\omega^{2}Nr^{4}-4r_{0}^{2}r^{2}+5r_{0}^{4})\;\phi_{+}(r)=0 \;.
\end{equation}
This equation admits the following solution:
\begin{equation}
 \label{042fermionmasslessmode}
\phi_{+}(r)=\sqrt{r}\;(r^{2}-r_{0}^{2})^{-1/4}\left(C_{1}\;(r^{2}-r_{0}^{2})^{-\frac{i}{2}\omega\sqrt{N}}
+\frac{C_{2}}{2i\omega\sqrt{N}}\;(r^{2}-r_{0}^{2})^{\frac{i}{2}\omega\sqrt{N}}\right) \;,
\end{equation}
where $C_{1}$ and $C_{2}$ are arbitrary constants.\par

Next, following the matching recipe \cite{Maldacena:1996ix, Gubser:1996zp, Klebanov:1996gy, Klebanov:1997cx}, we must match 
both solutions at asymptotic infinity and at near the event horizon in a matching point defined as $r_{m}$.
We must calculate the flux defined as
\begin{equation}
 \label{042flux}
{\cal F}=\frac{1}{2i}\left(\phi_{+}^{*}(r)r^{3}f(r)\partial_{r}\phi_{+}(r)-c.c.\right) \;.
\end{equation}
The ratio of the flux evaluated at the near horizon and at the 
asymptotic infinity, respectively, is the absorption 
cross section, which can be demonstrated that, in the low-energy regime, it is equal to the greybody factor. 
The mode solution at the near horizon limit is obtained imposing 
the propagation of ingoing modes as a boundary condition. Then, if we expand the solution (\ref{042fermionmasslessmode}) 
near the horizon, we obtain
\begin{equation}
 \label{042wavehorizon}
\phi_{h}(r)=C_{h}\;(r-r_{0})^{-\frac{1}{4}-\frac{i}{2}\omega\sqrt{N}} \;,
\end{equation}
where we have collected all the terms that are independent of the $r$ coordinate in the constant $C_{h}$. The corresponding
flux is
\begin{equation}
 \label{042fluxhorizon}
{\cal F}_{h}=\frac{\lvert C_{h}\lvert^{2}}{2}\;\omega\sqrt{N}\;\frac{r\;(r+r_{0})}{\sqrt{r-r_{0}}} \;.
\end{equation}
Next, we calculate the mode solution at the asymptotic limit. We must take into account that in this limit the metric 
function $f(r)$ fulfills the relation 
\begin{equation}
 \label{042finfinity}
\lim_{r \to \infty}f(r)=1 \;.
\end{equation}
Solving Eq. (\ref{042dirac2d3}) for the massless case using (\ref{042finfinity}), we obtain the following 
mode solution at the asymptotic limit:
\begin{equation}
 \label{042waveasymptotic}
\phi_{\infty}(r)=C_{\infty}\;r^{i\omega\sqrt{N}} \;.
\end{equation}
Therefore, the corresponding flux is
\begin{equation}
 \label{042fluxaymptotic}
{\cal F}_{\infty}=\frac{\lvert C_{\infty}\lvert^{2}}{2}\;\omega\sqrt{N}\;r^{2} \;.
\end{equation}
In order to find a relation between the constants $C_{h}$ and $C_{\infty}$, we match both solutions at the matching point 
$r_{m}$, which fulfills $r_{0}<<r_{m}$. Hence, imposing the matching condition, 
we find the following relation between the constants: 
\begin{equation}
 \label{matching}
\lvert C_{\infty}\lvert^{2}=\frac{\lvert C_{h}\lvert^{2}}{\sqrt{r_{m}-r_{0}}} \;.
\end{equation}
Finally, if we calculate the greybody factor as the ratio of the ingoing flux through the horizon, ${\cal F}_{h}$, 
to the outgoing flux at the asymptotic limit, ${\cal F}_{\infty}$, we obtain
\begin{equation}
 \label{cross}
\Gamma_{\omega}\equiv\frac{\lvert {\cal F}_{h}\lvert}{\lvert {\cal F}_{\infty}\lvert}=1 \;.
\end{equation}
This result indicates that for LST we obtain a pure Planckian spectrum. 
Effectively, one would expect this result since we have demonstrated in \cite{LorenteEspin:2007gz,LorenteEspin:2011pu} 
how LST exhibits a purely thermal behavior, even taking into account the backreaction of the metric.

\section{Conclusion}
In previous works the emission of massless scalar particles in the LST background had been studied; however, it is 
interesting 
to extend this study to the emission of Dirac particles in order to complete the emission description in LST.\par
We have calculated the emission of a fermion field in a two-dimensional effective LST metric relevant for the emission 
process. Interestingly, we also have found that the emission shows a purely thermal profile as in the scalar case. This 
lead us to the conclusion that the thermal profile of LST is independent of the emitted fields; the independent 
mass temperature (\ref{temp}) is the clue for this behavior.\par  
The gravitational potential barrier around the black hole acts as a filter for the emitted radiation, therefore the 
spectrum detected at the asymptotic infinity is not a pure Planckian spectrum. The greybody factor accounts for this 
deviation from the purely blackbody spectrum.
However we have verified that LST exhibits a different behavior; 
the independence of its temperature on the black hole mass leads to the fact that the 
emission is purely thermal, even taking into account backreaction effects. Therefore, one expects that the spectrum shall 
be purely Planckian and the greybody factor takes the value 1. We have verified this last assumption both for the emission 
of scalar particles and the emission of fermions in a two-dimensional effective metric.\par
Even if one considers backreaction and self-gravitating effects, the flux (\ref{034flux}) remains fixed at the Hagedorn 
temperature (\ref{temp}). Thus, the results obtained in this work, corresponding to the emission of fermions in LST 
background, match the same conclusions concerning the information loss paradox in scalar field emission studied in 
Refs. \cite{LorenteEspin:2007gz,LorenteEspin:2011pu}. Hence, this work completes the study concerning the emission of 
different kind of fields in LST.  

\vskip10mm
\noindent{\bf Acknowledgments:}\\
I thank Pere Talavera for discussions on this work.

\end{document}